\documentstyle[twoside,fleqn,espcrc2]{article}

\newcommand{\AmS}{{\protect\the\textfont2
  A\kern-.1667em\lower.5ex\hbox{M}\kern-.125emS}}

\title{Glueball Masses in Quantum Chromodynamics
}

\author{Xiang-Qian Luo, Qizhou Chen, Shuohong Guo,
Xiyan Fang, Jinming Liu\address{CCAST
     (World Laboratory),  P.O. Box 8730, Beijing 100080, China,\\
       Department of Physics,
       Zhongshan University, Guangzhou 510275, China,\\
       and Center for Computational Physics,
       Zhongshan University, Guangzhou 510275, China
             }
                          }

\begin{document}

\begin{abstract}
We review the recent glueball mass calculations 
using an efficient method for
solving the Schr\"odinger equation order by order
with a scheme preserving the continuum limit. 
The reliability of the method is further supported by
new accurate results 
for (1+1)-dimensional $\sigma$ models and (2+1)-dimensional
non-abelian models. 
We present first and encouraging data for the glueball
masses in 3+1 dimensional QCD. 
\end{abstract}

\maketitle

\section{INTRODUCTION}

Although lattice field theory is expected to give 
the most reliable estimates for QCD spectroscopy, 
the values for glueball masses have still been
an issue under debate \cite{Wein} for the last two decades.
Recently, more accurate numerical calculations 
by the IBM group \cite{IBM}
on much larger lattices and higher statistics lead to
$M(0^{++})\approx 1.740 ~Gev$ in the infinite volume 
and continuum limit. 

Alternatively,  
the lattice Hamiltonian methods \cite{Green,GCL,QCD3,MASS,SCHU,KS}
receive increasing attention, 
since more physical insights can be obtained and significant progress 
has been achieved in recent years.
The gluon dynamics is described by the Hamiltonian
\begin{eqnarray}
H= {g^{2} \over 2a} \sum_{l} E_{l} E_{l}
- {1
\over ag^{2}} \sum_{p} Tr(U_{p} + U_{p}^{\dagger} -2).
\label{H}
\end{eqnarray}
The most direct way 
to extract both the glueball masses and their wavefunctions 
is to solve the Schr{\"o}dinger eigenvalue equation 
$H \vert F \rangle = \epsilon_{F} \vert F \rangle$.
The glueball wavefunction $\vert F \rangle$ is
created by the gluonic operators
with the given  
$J^{PC}$ acting on the vacuum $\vert \Omega \rangle$.
The vacuum wavefunction $\vert \Omega \rangle$ satisfies 
$H \vert \Omega \rangle = \epsilon_{\Omega} \vert \Omega \rangle$.
An estimate of the glueball mass is then 
$M_{J^{PC}}=\Delta \epsilon=\epsilon_{F}-\epsilon_{\Omega}$.

Physically,
the low energy spectrum originates mainly 
from the long wavelength excitations.
In a series of papers \cite{GCL,QCD3,MASS}, 
we developed further a method for solving 
the eigenvalue problem \cite{Green}
in a new scheme preserving the correct long wavelength behavior 
at any order of approximation.
Having such a correct truncation is essential for
getting a correct continuum behavior of the physical quantities.
This idea and the scheme are confirmed by 
the results of 
three-dimensional abelian in \cite{FLG}
and non-abelian in \cite{GCL,QCD3,MASS,CGZF}
gauge theories: they converge rapidly, and even 
at very low truncation orders clear scaling
windows for the vacuum wavefunction and mass gaps
have been established. 
For 2+1 D $\rm{U(1)}$ and 2+1 D $\rm{SU(2)}$, 
they agree perfectly
with recent accurate
Monte Carlo data.
In our pioneering study of 2+1 D \rm{SU(3)}, 
we obtained the first estimates for the
vacuum wavefunction \cite{QCD3} and the glueball masses \cite{MASS}.
Most of these results and detailed techniques
have been summarized last year \cite{Guo,Luo}.

There have been new developments this year. 
Increasing evidences, higher order and more accurate data, 
including those \cite{SIGMA} 
for the two-dimensional $O(N)$ $\sigma$ models,
have been obtained to support the reliability
of the approach.
Here we would like to highlight our recent results \cite{LC,Hu}
for QCD in 2+1 and 3+1 dimensions.

\section{$\rm{QCD_3}$: BEYOND A TOY MODEL}

The reader will amaze later at the fact that 
$\rm{QCD_3}$ is not just a toy model of $\rm{QCD}_4$, but
it can even mimic the realistic theory. 
Using the techniques in \cite{MASS,COM} and 
analysing carefully the data in the observed scaling region
$\beta \in [5,12)$, a
more accurate value \cite{LC} for $M(0^{++})/e^2$ is obtained:  
\begin{eqnarray}
  {M(0^{++}) \over e^2} \approx 2.15 \pm 0.06,
\label{data1}
\end{eqnarray}
with the error being the systematic uncertainties due to finite
order truncation (in \cite{MASS},  
it was $2.1$
for narrower $\beta$ range at third order). 
(\ref{data1}) can be compared with 
Samuel's result $1.84 \pm 0.46$ in the continuum Hamiltonian  
formulation \cite{Samuel2} or 
the preliminary MC result 
$2.4 \pm 0.2$ on a finte lattice obtained much later \cite{TP}.

A relation between the glueball mass
and the confinement scale from the vacuum wavefunction may be induced.
The vacuum functional, which interpolates the strong and weak coupling
regimes, are \cite{Arisue,Samuel2}
\begin{eqnarray*}
\vert \Omega \rangle=exp \{ {1 \over 2 e^2} \int d^{D-1}x ~ 
tr [{\cal F} 
({\cal D}^2+ \xi^{-2})^{-1/2} {\cal F}] \}.
\end{eqnarray*}
The correlation length $\xi$, with dimension of inverse mass, 
is  proportional to $e^{-2}$, i.e.,
the confinement scale in the vacuum.  
$\xi^{-1}$ 
might also be relevant for the constituent gluon mass 
and the lightest glueball mass \cite{Samuel2}.
In the strong coupling or large $N_c$
limit, $\vert \Omega \rangle$ reduces to the strong coupling 
wavefunction \cite{Green,Feynman}.
In the intermediate and weak coupling, 
it becomes \cite{Arisue,GCL,QCD3}
\begin{eqnarray*}
\vert \Omega \rangle=exp \{ 
\int d^{D-1} x ~ \lbrack - {\mu_0} tr {\cal F}^2
- {\mu_2} ~tr ({\cal D} {\cal F})^2 \rbrack \},
\label{a1}
\end{eqnarray*}
identical to our long wavelength vacuum wavefunction \cite{GCL,QCD3}.
The correlation length is then related to $\mu_0$ and $\mu_2$ by 
$\xi=({-2 \mu_2 / \mu_0})^{1/2}$.
For 2+1 D SU(2), $\xi=0.65/e^2$ (see \cite{Arisue,GCL}),
while for 2+1 D SU(3), our result \cite{QCD3} is $\xi=0.53/e^2$. 
If the glueball mass is proportional to the constituent gluon mass,
from the difference of the scales between SU(2) and SU(3), one 
may also guess $M(0^{++})/e^2 \approx 2$, consistent with
(\ref{data1}) from our practical calculation.

Combining the recent MC data \cite{Lut} 
for the string tension 
$\sigma$ in $\rm{QCD}_3$, 
we obtain in the continuum limit
\begin{eqnarray}
  {M_{0^{++}} \over \sqrt{\sigma}} \approx 3.88 \pm 0.11.
\label{data2}
\end{eqnarray}

To extend $\rm{QCD_3}$ to $\rm{QCD_4}$, 
we follow the dimensional reduction argument \cite{Green,Samuel2},
which says that a confining theory in $D$ dimensions ($2 < D \leq 4$)
becomes a localized field theory in $d=D-1$ dimensions. 
This can be exactly proven in the strong coupling or large $N_c$ limit.
In this limit, the fixed time vacuum expectation value of an
operator $O(U)$ in $D$ dimensions 
corresponds to the path integral expression
for $<O(U)>$ in $D-1$ dimensional lattice field theory.
In the intermediate coupling
region the 3+1 D theory can still 
be approximated by its 2+1 D theory
for long wavelength configurations 
in comparison to the confinement scale. Accordingly, 
$M(J^{PC})/\sqrt{\sigma}$ for the lightest glueball 
should be approximately the same
for 2+1 and 3+1 dimensions. 
Since for SU(3), $N_c$ is larger and 
the measured length $\xi$ in the vacuum functional ({\ref{a1})  
is smaller than those for SU(2), our speculation is 
that dimensional reduction work better. 
In fact, (\ref{data2}) is well consistent
with the IBM data $M(0^{++})/\sqrt{\sigma}=3.95$ from
MC simulation of $\rm{QCD_4}$, providing 
the continuum $\sqrt{\sigma}=0.44$ Gev is used.
From this world averaged $\sqrt{\sigma}$ and (\ref{data2}),
we expect 
\begin{eqnarray}
M(0^{++})=1.71 \pm 0.05 ~ GeV, 
\end{eqnarray}
in nice agreement with the IBM
data $M(0^{++})=1.740 \pm 0.071$ \cite{IBM}.
This favors $\theta /f_J(1710)$ as a candidate of the $0^{++}$ glueball.

\section{$\rm{QCD_4}$: THE REALISTIC THEORY}

It is very desirable to do concrete computations 
in the realistic theory: \rm{QCD} in 3+1 dimensions.
In \cite{Hu}, we made a first step towards this direction.
As our first attempt,
we computed the masses of glueballs $0^{++}$, $0^{--}$, and $1^{+-}$,
which gluonic operators are easily constructed. 

At relatively strong coupling, the absolute value of $aM_E$, 
calculated in the Hamiltonian formulation and converted to
the Euclidean one, differs
from the result from Euclidean strong coupling expansion. 
This is not surprising because
they are different schemes 
at finite lattice spacing,
and the weak coupling relation between them doesn't not hold
for strong coupling. 
For $\beta \ge 6.0$, we do observe much smaller difference in $aM_E$ 
and a tendency 
approaching the MC data \cite{Wein}. 

Similar to the most recent MC data \cite{Wein} 
in the available coupling region,
clear asymptotic scaling window for the individual mass 
$aM(0^{++})$, $aM(0^{--})$, or $aM(1^{+-})$ could not be found. 
There might be two possible reasons for this scaling violation:
the available coupling region being not weak enough for 
the asymptotic scaling law 
to be valid or the results being not accurate enough.
The first one may be reduced by the Symanzik's improvement or
the Lepage-Mackenzie scheme. The second one
may be improved by higher order calculations.
For the mass ratio $M_E^{(1)}/M_E^{(2)}$, however, 
the large part of errors in $M_E^{(1)}$ and $M_E^{(2)}$ are
cancelled. Indeed, we observe approximate plateau in the mass ratio 
both for $M(0^{--})/M(0^{++})$ and $M(1^{+-})/M(0^{++})$
in the coupling region $\beta \in [6.0,6.4]$.

From the plateaus in $\beta \in [6.0,6.4]$,
we estimate 
\begin{eqnarray*}
{M(0^{--}) \over M(0^{++})}=2.44 \pm 0.05 \pm 0.20,
\end{eqnarray*}
\begin{eqnarray}
{M(1^{+-}) \over M(0^{++})}=1.91 \pm 0.05 \pm 0.12,
\label{data}
\end{eqnarray}
where 
the mean value is the averaged one over the data in this region,
the first error is the error of the data in the plateau, and
the second error is the upper limit of the systematic errors
due to the finite order truncation.

For comparison, we list the corresponding
results from other lattice calculations.
For $M(0^{--})/M(0^{++})$, it is approximately 
$3.43 \pm 1.50$ from Monte Carlo simulations with a huge error bar 
(see the references in \cite{Wein}), and $2.2 \pm 0.2$ from
t-expansion plus Pad\'e approximants \cite{Horn}.
For $M(1^{+-})/M(0^{++})$, it is $1.88 \pm 0.02$ from Monte Carlo,
and $1.60 \pm 0.60$ from t-expansion plus Pad\'e.
For $M(0^{--})/M(0^{++})$, we obtain a value between the 
Monte Carlo and t-expansion data, with error under much better control than
the former one. For the $M(1^{+-})/M(0^{++})$, where the systematic
error in the MC data is very small, we get a value 
consistent well with them.

To summarize, we have tested with higher precision our new scheme 
\cite{GCL,QCD3,MASS} 
in 1+1 D $\sigma$ models and 2+1 D
non-abelian models. 
For the lightest
glueball mass in $\rm{QCD_3}$, 
accurate results are obtained with systematic uncertainty under well control.
The idea of dimensional reduction is successively applied to extrapolate 
$\rm{QCD_3}$ to $\rm{QCD_4}$. 
We have also done concrete calculations of QCD in 3+1 D and
presented first and encouraging data for some glueball
masses.
The inclusion of other glueballs, such as 
$2^{++}$, $0^{-+}$ and $0^{+-}$, and reduction of the systematic errors
are in progress.

Our work was sponsored by 
the National Natural Science Foundation
and Education Committee.
We are grateful to C. Bernard, J. Greensite, H. Kr\"oger,
S. Samuel, N. Scheu, D. Sch{\"u}tte  
and BES members for useful discussions. 
XQL thanks the Lattice 96 organizing committee for
support.


\begin{thebibliography}{9}

\bibitem{Wein} D.Weingarten, Nucl.Phys.{\bf B(PS)34}(1993)29.

\bibitem{IBM} J. Sexton, A. Vaccarino, and D. Weingarten, 
Phys. Rev. Lett. {\bf 75} (1995) 4563.

\bibitem{Green} J. Greensite,
Nucl. Phys. B{\bf 166} (1980) 113.

\bibitem{GCL} S.Guo,Q.Chen,L.Li,Phys.Rev.{\bf D49}(1994)507.

\bibitem{QCD3} Q. Chen, X.Q. Luo, and S. Guo,
 Phys. Lett. B{\bf 341} (1995) 349.

\bibitem{MASS} Q.Z. Chen, X.Q. Luo, S.H. Guo and X.Y. Fang,
Phys. Lett. {\bf B348} (1995) 560.

\bibitem{SCHU} C. Hamer, W. Zheng and D. Sch\"utte,
hep-ph/9511179; hep-lat/9603026.

\bibitem{KS} H. Kr\"oger and N. Scheu, 
hep-lat/9508007; hep-lat/9607006; hep-lat/9607008.

\bibitem{FLG} X.Fang,J.Liu,S.Guo, Phys. Rev. {\bf D53}(1996).

\bibitem{CGZF} Q. Chen, S. Guo, W. Zheng
 and X. Fang, Phys. Rev. D{\bf 50} (1994) 3564;
Zhongshan Preprint (1995).

\bibitem{Guo} S. Guo, Q. Chen, X. Fang, J. Liu, X.Q. Luo 
and W. Zheng, Nucl. Phys. {\bf B(Proc. Suppl.)47} (1996) 827.

\bibitem{Luo} Q.Z. Chen, S.H. Guo, X.Q. Luo and A. Segu\'i, 
Nucl. Phys. {\bf B(Proc. Suppl.)47} (1996) 274.

\bibitem{SIGMA} S.H. Guo et al., Zhongshan Preprint (1996).

\bibitem{LC} X.Q. Luo and Q.Z. Chen, hep-ph/9604395.

\bibitem{Hu} L. Hu, X.Q. Luo, Q. Chen, X. Fang, and S. Guo,
Zhongshan Preprint (1996).

\bibitem{COM} X. Luo, Q.Chen,P.Cai and S.Guo, to appear.

\bibitem{Samuel2} S. Samuel, hep-ph/9604405.

\bibitem{TP} M. Teper, these proceedings.

\bibitem{Arisue} H. Arisue, Phys. Lett. B{\bf 280} (1992) 85.

\bibitem{Feynman} R. Feynman, Nucl. Phys. {\bf B188} (1981) 479.

\bibitem{Lut} M. L\"utgemeier, 
Nucl. Phys. {\bf B (Proc. Suppl.)42} (1995) 523; these proceedings.

\bibitem{Horn} C. van den Doel and D. Horn, 
Phys. Rev. {\bf D35} (1987) 2824.

\end{thebibliography}
\end{document}